\documentclass[12pt]{article}
\usepackage{epsfig,amsfonts,amsthm}
\usepackage{amsmath,amssymb}
\usepackage{cite}
\usepackage{xcolor}
\usepackage{tikz}
\usepackage{extarrows}
\usepackage{url}

\newcommand{\be}{\begin{equation}}
\newcommand{\ee}{\end{equation}}
\newcommand{\bea}{\begin{eqnarray}}
\newcommand{\eea}{\end{eqnarray}}

\newcommand{\triplet}[3]{ \left(\! \begin{array}{c}#1 \\ #2 \\ #3 \end{array}\!\right) }
\providecommand{\mtrx}[1]{\begin{pmatrix} #1 \end{pmatrix}}

\newcommand{\Z}{\mathbb{Z}}
\newcommand{\mmmatrix}[9]{ \left(\!\! \begin{array}{ccc}#1 & #2 & #3\\ #4 & #5 & #6\\ #7 & #8 & #9\\ \end{array}\!\!\right) }
\providecommand{\id}{{\boldsymbol{1}}}

\def\lsim{\mathrel{\rlap{\lower4pt\hbox{\hskip1pt$\sim$}}
    \raise1pt\hbox{$<$}}}         
\def\gsim{\mathrel{\rlap{\lower4pt\hbox{\hskip1pt$\sim$}}
    \raise1pt\hbox{$>$}}}         

\title{3HDM with softly broken $\Delta(54)$ and $\Sigma(36)$}

\author{Gonçalo~Barreto$^{1}$\thanks{E-mail: goncalormbarreto@tecnico.ulisboa.pt},	
	Ivo~de~Medeiros~Varzielas$^{1}$\thanks{E-mail: ivo.de@udo.edu}  
	\\
	{\small $^1$ CFTP, 
		Instituto Superior T\'{e}cnico, Universidade de Lisboa, 
		av. Rovisco Pais 1, 1049-001 Lisboa, Portugal}
}

\begin{document}
\maketitle

	\bigskip
\begin{abstract}
We perform an analysis of the scalar sector of 3 Higgs Doublet Models with softly broken $\Delta(54)$ and $\Sigma(36)$ symmetries.
We consider the various vacuum expectation value alignments and consider, for each, softly broken terms that deviate the alignment. We check the evolution of the minima, present analytical and numerical results for the lifting of the degeneracies of the physical eigenstates, and describe the decays of the states considering any residual symmetries.

\end{abstract}


\section{Introduction}

The Standard Model (SM) is extremely successful but there are several indications that the SM is not the complete theory. Among the many possibilities of Beyond Standard Model (BSM) theories, the simple idea of having more than one Higgs $SU(2)$ doublet is well motivated, because it is a framework that can readily provide Dark Matter (DM) candidates and enable the possibility of spontaneous CP violation. For recent reviews, see e.g. \cite{Branco:2011iw,Kanemura:2014bqa,Ivanov:2017dad,Arcadi:2019lka}.

The potential of the most general Higgs Doublet Model can be written as:
\begin{equation}
V=Y_{ij} (\phi^{\dagger}_i\phi_j)+Z_{ijkl} (\phi^{\dagger}_i\phi_j)(\phi^{\dagger}_k\phi_l)\,, \quad i,j,k,l = 1, \dots, N,\label{YZ}
\end{equation}
up to renormalisable level, and where $N$ is the number of doublets.
Focusing on the case of 3 Higgs Doublet Models (3HDM), this expression has 54 free parameters.

The list of discrete symmetries for the 3HDM that don't lead to a renormalisable potential accidentally invariant under a continuous symmetry is small \cite{Ivanov:2012fp}. \footnote{See also \cite{Darvishi:2019dbh,Darvishi:2021txa}.}
Considering the symmetries with a triplet irreducible representation, there is $A_4$, $S_4$, $\Delta(54)$ and $\Sigma(36)$, the number of free parameters in the respective potential is greatly reduced:
\begin{equation}
V = - m^2 (\phi_1^\dagger \phi_1 + \phi_2^\dagger \phi_2 + \phi_3^\dagger \phi_3) + V_4\,,
\end{equation}
where $V_4$ depends on the specific symmetry, and the other part is common to the 4 symmetries.
The respective minima have been found using different methods
\cite{Ivanov:2014doa,deMedeirosVarzielas:2017glw}.

For each minimum, the masses (and respective degeneracies) of the physical states can be calculated. Further, when softly breaking the potential, the soft breaking parameters (SBPs) are in general:
\begin{eqnarray}
V_{\rm soft} = m_{11}^2 \phi_1^\dagger\phi_1 + m_{22}^2 \phi_2^\dagger\phi_2 + m_{33}^2 \phi_3^\dagger\phi_3 
+ \left(m_{12}^2\,\phi_1^\dagger\phi_2 + m_{23}^2\,\phi_2^\dagger\phi_3 
+ m_{31}^2\,\phi_3^\dagger\phi_1 + h.c.\right),\label{soft}
\end{eqnarray}
with complex $m_{ij}^2$ for $i \not = j$, accounting for 9 free parameters.
They can be usefully classified as alignment-preserving (direction of the minima remains unchanged) or otherwise. The classification was suggested in \cite{deMedeirosVarzielas:2021zqs}, together with the example for the $\Sigma(36)$ case.  Analogously, the symmetric limit and the softly broken $A_4$, $S_4$ potential was analysed in \cite{deMedeirosVarzielas:2022kbj}, finding for some of the minima cases with residual symmetries unbroken by the minima and by the alignment-preserving SBPs. These residual symmetries can stabilize physical states preventing their decay.

In this work, we consider both the softly broken $\Delta(54)$ and $\Sigma(36)$ cases, with SBPs that do not preserve the direction of the minima of the symmetric limit (as noted above, the alignment-preserving case has been previously studied for $\Sigma(36)$ in particular \cite{deMedeirosVarzielas:2021zqs}). Accordingly, we consider the effect of the soft breaking on the direction of the minima, on the mass eigenstates and respective masses.

The layout of the paper is as follows. In Section \ref{section: exact Sigma36} we look at the $\Sigma(36)$ model and present the results for the softly broken case; in Section \ref{section: exact Delta54} we show the results for the $\Delta(54)$ potential; in Section \ref{section: decays} the decays of both models are studied; the conclusions are presented in Section \ref{section: conclusion}.


\section{$\Sigma(36)$-symmetric 3HDM}\label{section: exact Sigma36}

\subsection{The scalar potential and its minima}

The $\Sigma(36)$ and the $\Delta(54)$ are the largest discrete symmetry groups that can be imposed on the scalar sector of 3HDM that do not lead to accidental continuous symmetries \cite{Ivanov:2012fp}.
The group $\Sigma(36)$ is defined as a $\Z_4$ permutation acting on generators of the abelian group $\Z_3\times \Z_3$:

\begin{equation}
\Sigma(36) \simeq (\Z_3 \times \Z_3)\rtimes \Z_4\,.
\end{equation}

The generators of both the $\Z_3$ groups and the generator of $\Z_4$ are, correspondingly:

\begin{equation}
a = \mtrx{1&0&0\\ 0&\omega&0\\ 0&0&\omega^2}\,, \quad
b = \mtrx{0&1&0\\ 0&0&1\\ 1&0&0}\,,\quad
d = {i \over\sqrt{3}} \left(\begin{array}{ccc} 1 & 1 & 1 \\ 1 & \omega^2 & \omega \\ 1 & \omega & \omega^2 \end{array}\right)\,,
\label{eq: Sigma36-generators}
\end{equation}

where $\omega = \exp(2\pi i/3)$. These generators have the following orders:
\[
a^3 = \id\,,\quad b^3 = \id\,, \quad d^4 = \id\,.
\]

The scalar potential of 3HDM invariant under $\Sigma(36)$ is the following:
\begin{eqnarray}
V_0 & = &  - m^2 \left[\phi_1^\dagger \phi_1+ \phi_2^\dagger \phi_2+\phi_3^\dagger \phi_3\right]
+ \lambda_1 \left[\phi_1^\dagger \phi_1+ \phi_2^\dagger \phi_2+\phi_3^\dagger \phi_3\right]^2 \nonumber\\
&&
- \lambda_2 \left[|\phi_1^\dagger \phi_2|^2 + |\phi_2^\dagger \phi_3|^2 + |\phi_3^\dagger \phi_1|^2
- (\phi_1^\dagger \phi_1)(\phi_2^\dagger \phi_2) - (\phi_2^\dagger \phi_2)(\phi_3^\dagger \phi_3)
- (\phi_3^\dagger \phi_3)(\phi_1^\dagger \phi_1)\right] \nonumber\\
&&
+ \lambda_3 \left(
|\phi_1^\dagger \phi_2 - \phi_2^\dagger \phi_3|^2 +
|\phi_2^\dagger \phi_3 - \phi_3^\dagger \phi_1|^2 +
|\phi_3^\dagger \phi_1 - \phi_1^\dagger \phi_2	|^2\right)\, .
\label{Vexact}
\end{eqnarray}

By using geometric minimization \cite{Degee:2012sk}, one can show that both the $\Sigma(36)$ and $\Delta(54)$ 3HDM minima always has the following radial directions:

\begin{eqnarray}
\mbox{Alignment $A$:}&& A_1 = (\omega,\,1,\,1)\,, \quad A_2 = (1,\,\omega,\,1),\, \quad A_3 = (1,\,1,\,\omega)\label{points-A} \nonumber\\
\mbox{Alignment $A'$:}&& A'_1 = (\omega^2,\,1,\,1)\,, \quad A'_2 = (1,\,\omega^2,\,1),\, \quad A'_3 = (1,\,1,\,\omega^2)\label{points-Ap} \nonumber\\
\mbox{Alignment $B$:}&& B_1 = (1,\,0,\,0)\,, \quad B_2 = (0,\,1,\,0),\, \quad B_3 = (0,\,0,\,1)\label{points-B} \nonumber\\
\mbox{Alignment $C$:}&& C_1 = (1,\,1,\,1)\,,\quad C_2 = (1,\,\omega,\,\omega^2)\,,\quad C_3 = (1,\,\omega^2,\,\omega)\label{eq: alignments}
\end{eqnarray}

The potential has four real free parameters. Depending on the relations between these parameters the true minima will belong to a different alignment. The conditions for the selection of each alignment are as follows:

\begin{eqnarray}
\mbox{Alignments $A$ + $A'$:}&& \lambda_3 < 0 \nonumber\\
\mbox{Alignment $B$:}&& \lambda_3 > 0 \nonumber\\
\mbox{Alignment $C$:}&& \lambda_3 > 0\label{eq: alignment selection Sigma 36}
\end{eqnarray}

Since the potential conserves CP, the alignments $A$ and $A'$ merge.
In the $\Sigma(36)$ case, the alignments $B$ and $C$ merge, as they are related by the $d$ transformation.
However, for consistency with the $\Delta(54)$ case, they are presented separately.

\subsection{The physical Higgs bosons}

Three complex Higgs doublets contain 12 real fields. When expanding the potential around a neutral vacuum, one absorbs, as usual, three of them in the longitudinal components of
the $W^\pm$ and $Z$-bosons. What remains is two pairs of charged Higgses and five neutral Higgs bosons.
At points $B$ and $C$, the Higgs boson masses are:

\begin{eqnarray}
m_{h_{SM}}^2 &=& 2m^2\,, \nonumber\\[1mm]
m_{H^\pm}^2 &=& \lambda_2 v^2\quad \mbox{(double degenerate)}\,,\nonumber\\[1mm]
m_{h}^2 &=& \lambda_3 v^2\quad \mbox{(double degenerate)}\,,\nonumber\\[1mm]
m_{H}^2 &=& 3\lambda_3 v^2\quad \mbox{(double degenerate)}\,.\label{Sigma 36-masses BC}
\end{eqnarray}

At points $A$ and $A'$, the masses are as follows:

\begin{eqnarray}
m_{h_{SM}}^2 &=& 2m^2\,, \nonumber\\[1mm]
m_{H^\pm}^2 &=& (\lambda_2 - 3\lambda_3) v^2\quad \mbox{(double degenerate)}\,,\nonumber\\[1mm]
m_{h}^2 &=& -\lambda_3 v^2\quad \mbox{(double degenerate)}\,,\nonumber\\[1mm]
m_{H}^2 &=& -3\lambda_3 v^2\quad \mbox{(double degenerate)}\,.\label{Sigma 36-masses A}
\end{eqnarray}

\subsection{Softly broken potential}

The discrete symmetry groups $\Delta(54)$ and $\Sigma(36)$ lead to a very strict phenomenology. The predictions made by these models can readily be in conflict with experiment (see e.g. \cite{Kalinowski:2021lvw}). It is therefore relevant to consider soft breaking parameters. In the case of 3HDM one can add up to 9 free parameters. SBPs are quadratic terms that are not invariant under the actions of the group. These parameters are usually considered to be parametrically small.

The goal of this work is to study how these soft breaking terms change the structural properties of the scalar sector of these 3HDM. As a simple example, we have added to the potential the following soft breaking matrix:

\begin{equation}
V_{\rm soft} = \phi_i^\dagger M_{ij} \phi_j\,\quad 
M_{ij} = \mmmatrix{0}{0}{0}%
{0}{m_{22}^2}{0}%
{0}{0}{0}\,,
\label{eq: SBM}
\end{equation}
where we choose, without loss of generality, one of the diagonal SBPs, which as we will see, changes the VEV alignment.

We now start by examining what happens to the VEV alignment (1,1,1). Any of the diagonal SBPs will not preserve the VEV alignment \cite{deMedeirosVarzielas:2021zqs}. Taking e.g. the $m_{22}^2$ term, the VEV will have the following form:

\begin{equation}
(v,v,v)/\sqrt{3}\, \xlongrightarrow{m_{22}^2} \,(v,u,v)/\sqrt{3}\,.
\end{equation}

The evolution of $v$ and $u$ as a function of $m_{22}^2$ is shown in Figure \ref{fig: uv Sigma36}.
The parameter $m_{22}^2$ was considered to be small compared to $m^2$.

\begin{figure}[ht]
    \centering
    \includegraphics[width=0.6\textwidth]{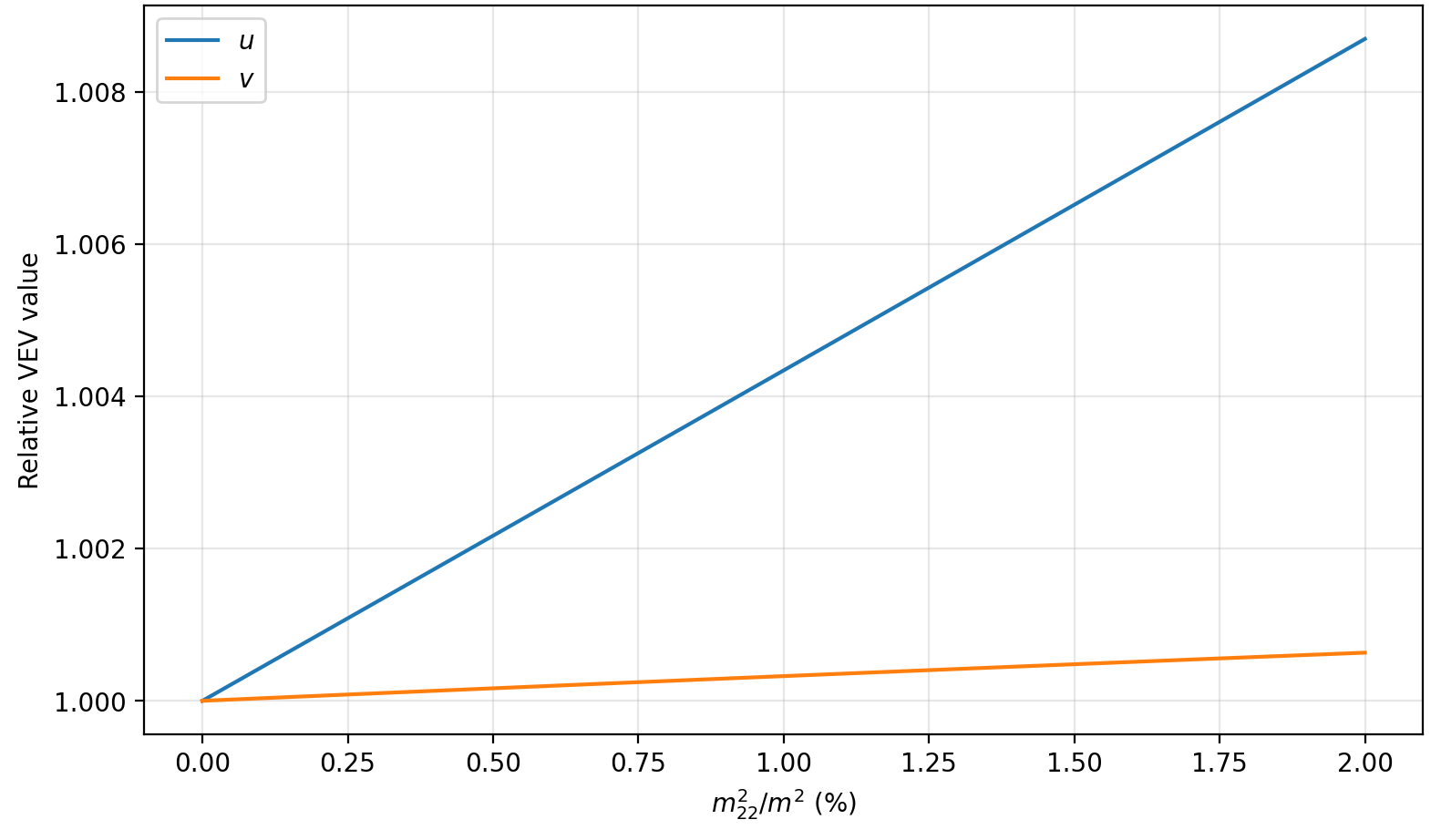}
    \caption{Evolution of the parameters $v/v_0$ and $u/v_0$ as a function of $m_{22}^2$ for $\Sigma(36)$. Where $v_0$ corresponds to $m_{22}^2=0$}
    \label{fig: uvSigma36}
\end{figure}

For the parameter space where (1,1,1) is a minima both $v$ and $u$ are positive real numbers and if $m_{22}^2>0$, then $u>v$. The masses of the physical fields were computed with the respective eigenvectors, in the basis of eq.(\ref{eq: Sigma36-generators}). The masses of the scalar fields are:

\begin{eqnarray}
m_{h_{SM}}^2 &=& m^2 - \frac{m_{22}^2}{2} - \frac{u^2}{2} \lambda_{3} + u v \lambda_{3} - \frac{1}{2}\sqrt{4 m^4 + m_{22}^4 - m_{22}^2 u^2 \lambda_{3} +
7 m_{22}^2 u v \lambda_{3}}\nonumber\\
&& \overline{+ v^2 (u^2- 2 u v + 2 v^2) \lambda_{3}^2 - 
\frac{32}{3} m_{22}^2 v^2 (2 \lambda_{1} + \lambda_{3}) +
4 m^2 (5 m_{22}^2 + u (u - 2 v) \lambda_{3})} ,\nonumber\\
m_{H1^\pm}^2 &=& -m_{22}^2 + \frac{u^2}{3} \lambda_{2} + \frac{2}{3} v^2 \lambda_{2} - \frac{2}{3}(u - v)^2 \lambda_{3} ,\nonumber\\
m_{H2^\pm}^2 &=& \frac{2}{3} v^2 (\lambda_{2} - 2 \lambda_{3}) +
\frac{u^2}{3} (\lambda_{2} - \lambda_{3}) + \frac{5}{3} u v \lambda_{3}
,\nonumber\\
m_{h1}^2 &=& m^2 - \frac{m_{22}^2}{2} - \frac{u^2}{2} \lambda_{3} + u v \lambda_{3} + \frac{1}{2} \sqrt{4 m^4 + m_{22}^4 - m_{22}^2 u^2 \lambda_{3} +
7 m_{22}^2 u v \lambda_{3}}\nonumber\\
&&
\overline{+ v^2 (u^2 - 2 u v + 2 v^2) \lambda_{3}^2 -
\frac{32}{3} m_{22}^2 v^2 (2 \lambda_{1} + \lambda_{3}) +
4 m^2 (5 m_{22}^2 + u (u - 2 v) \lambda_{3})},\nonumber\\
m_{h2}^2 &=& (u^2 + 5 u v - 4 v^2) \frac{\lambda_{3}}{3},\nonumber\\
m_{H1}^2 &=& -m_{22}^2 + (u + 2 v )^2 \frac{\lambda_{3}}{3},\nonumber\\
m_{H2}^2 &=& 3 u v \lambda_{3}.
\label{eq: Sigma 36-masses}
\end{eqnarray}

It is simple to verify that by setting $m_{22}^2 = 0$ and $u = v$, one obtains the masses in the exact case eq.(\ref{Sigma 36-masses BC}).

The value of the potential at the VEV is given by:

\begin{equation}
    V|_{VEV} = \frac{1}{6} (-m_{22}^2 u^2 - m^2 (u^2 + 2 v^2)) \,. 
\end{equation}

\subsection{Computational Example}

To better understand the results of the previous section we choose one example of parameters that gave us the following masses:

\begin{eqnarray}
    m_{h_{SM}} &=& 125.1 GeV\nonumber\\
    m_{H1^\pm} &=& 115.0 GeV\nonumber\\
    m_{H2^\pm} &=& 115.3 GeV\nonumber\\
    m_{h1} &=& 139.5 GeV\nonumber\\
    m_{h2} &=& 140.1 GeV\nonumber\\
    m_{H1} &=& 242.1 GeV\nonumber\\
    m_{H2} &=& 242.3 GeV\nonumber\\
\end{eqnarray}

The soft breaking term took away the degeneracies between several masses. For example the four charged bosons split into two pairs of charged bosons. The pairs of light and heavy scalars also get split.

\begin{figure}[ht]
    \centering
    \includegraphics[width=0.6\textwidth]{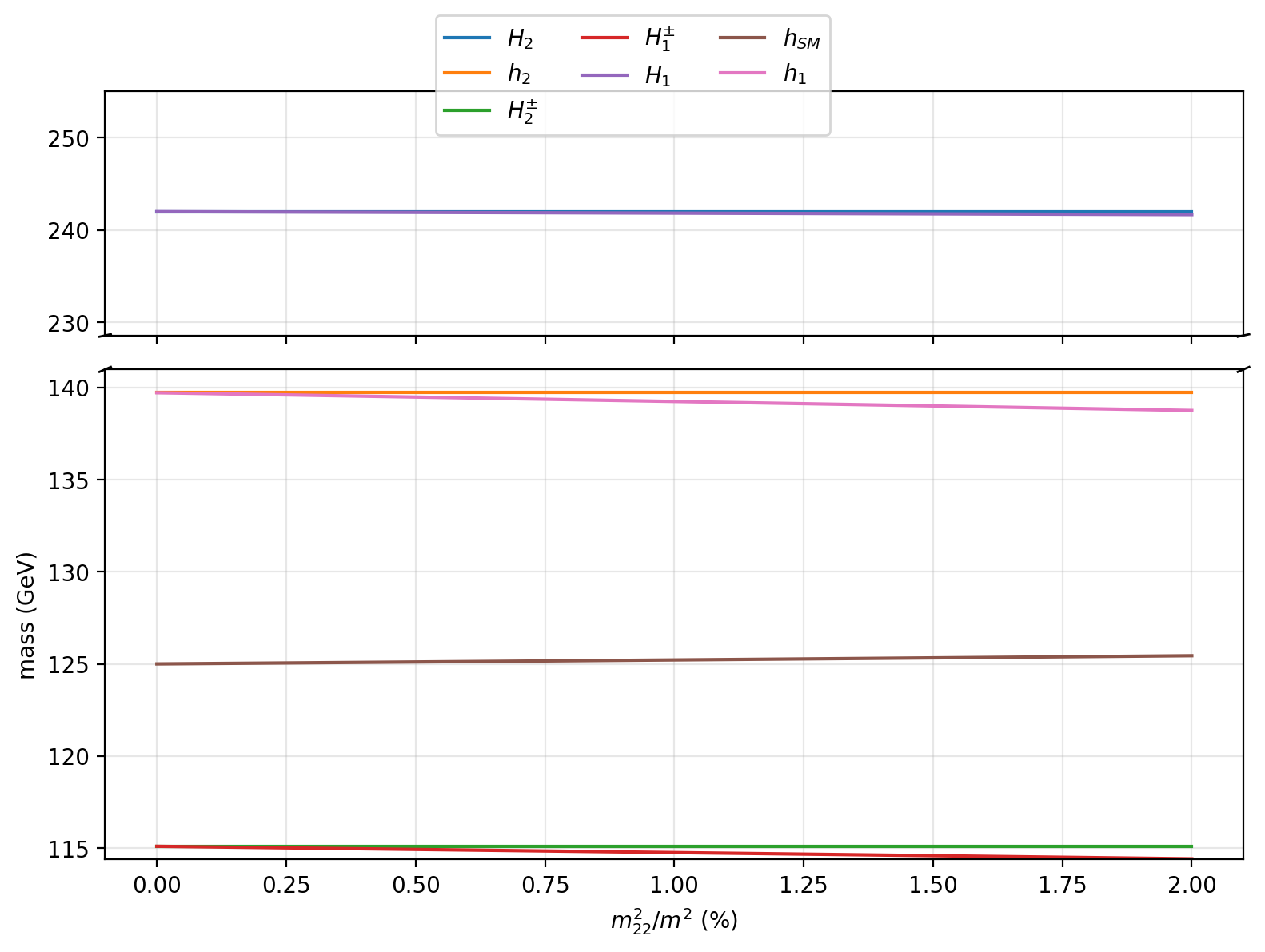}
    \caption{Masses evolution as a function of $m_{22}^2$ for $\Sigma(36)$.}
    \label{fig: Masses Sigma36}
\end{figure}

The evolution of the masses as a function of $m_{22}^2$ is shown in Figure \ref{fig: Masses Sigma36}.
The magnitude of the splitting varies between the different states.

All the cases with a single SBP turned on were systematically examined, which we do not explicitly present here as the results are similar to those shown previously. For example, when starting from the VEV $(1,\omega,1)$ and turning on $m_{22}^2$ or starting from the VEV $(1,\omega,1)$ and turning on the off-diagonal SBP $m_{12}^2$ instead of $m_{22}^2$, in both cases, the four charged boson masses are split into two pairs, and the masses of the light and heavy pairs of neutral scalars also get split.


\section{CP conserving $\Delta(54)$ 3HDM}\label{section: exact Delta54}

\subsection{The scalar potential and its minima}

In this section the general properties of the CP conserving $\Delta(54)$ 3HDM are presented. The group $\Delta(54)$ is defined as a $\Z_2$ permutation acting on generators of the Abelian group $\Z_3\times \Z_3$:

\begin{equation}
\Delta(54) \simeq (\Z_3 \times \Z_3)\rtimes \Z_2\,.
\end{equation}

The generators of the $\Delta(54)$ group are:

\begin{equation}
a = \mtrx{1&0&0\\ 0&\omega&0\\ 0&0&\omega^2}\,, \quad
b = \mtrx{0&1&0\\ 0&0&1\\ 1&0&0}\,,\quad
d^2 = \left(\begin{array}{ccc} -1 & 0 & 0 \\ 0 & 0 & -1 \\ 0 & -1 & 0 \end{array}\right)\,.
\label{eq: Delta 54-generators}
\end{equation}

These generators have the following orders:

\[
a^3 = \id\,,\quad b^3 = \id\,, \quad (d^2)^2 = \id\,.
\]

Notice that we imposed the symmetry under $d^2$ but not $d$. If we had imposed the symmetry under $d$ one would obtain the previous case of the $\Sigma(36)$.

The scalar potential of 3HDM invariant under CP conserving $\Delta(54)$ is the following \footnote{In the CP conserving $\Delta(54)$ potential the $\lambda_{4}$ term may have relative phases. In this paper these phases where considered 0 for simplicity. For more insight see \cite{deMedeirosVarzielas:2019rrp}.}:

\begin{eqnarray}
V_0 & = &  - m^2 \left[\phi_1^\dagger \phi_1+ \phi_2^\dagger \phi_2+\phi_3^\dagger \phi_3\right]
+ \lambda_1 \left[\phi_1^\dagger \phi_1+ \phi_2^\dagger \phi_2+\phi_3^\dagger \phi_3\right]^2 \nonumber\\
&&
- \lambda_2 \left[|\phi_1^\dagger \phi_2|^2 + |\phi_2^\dagger \phi_3|^2 + |\phi_3^\dagger \phi_1|^2
- (\phi_1^\dagger \phi_1)(\phi_2^\dagger \phi_2) - (\phi_2^\dagger \phi_2)(\phi_3^\dagger \phi_3)
- (\phi_3^\dagger \phi_3)(\phi_1^\dagger \phi_1)\right] \nonumber\\
&&
+ \lambda_3 \left(
|\phi_1^\dagger \phi_2 - \phi_2^\dagger \phi_3|^2 +
|\phi_2^\dagger \phi_3 - \phi_3^\dagger \phi_1|^2 +
|\phi_3^\dagger \phi_1 - \phi_1^\dagger \phi_2	|^2\right) \nonumber\\
&& 
+ \lambda_4 \left(
(\phi_1^\dagger \phi_3) (\phi_2^\dagger \phi_3) +
(\phi_2^\dagger \phi_1) (\phi_3^\dagger \phi_1) +
(\phi_3^\dagger \phi_2) (\phi_1^\dagger \phi_2) + h.c.	\right)\, 
\label{eq: V delta 54 exact}
\end{eqnarray}

The minima always belong to the alignments in eq.(\ref{eq: alignments}). The potential has five real free parameters. Depending on the relations between these parameters the true minima will belong to a different alignment. The conditions for the selection of each alignment are as follows:

\begin{eqnarray}
\mbox{Alignments $A$ + $A'$:}&& 9\lambda_1^2\lambda_3 + 5\lambda_3\lambda_4^2 + 9\lambda_1\lambda_3\lambda_4 + 3\lambda_1^2\lambda_4 < 2\lambda_4^3 + 3\lambda_1\lambda_4^2 + 4\lambda_3^2\lambda_4\,, \quad 3\lambda_3 < \lambda_4 \nonumber\\
\mbox{Alignment $B$:}&& 3\lambda_3 > \lambda_4\,, \quad \lambda_4 > 0 \nonumber\\
\mbox{Alignment $C$:}&& 9 \lambda_1^2\lambda_3 + 5\lambda_3\lambda_4^2 + 9\lambda_1\lambda_3\lambda_4 + 3\lambda_1^2\lambda_4 > 2\lambda_4^3 + 3\lambda_1\lambda_4^2 + 4\lambda_3^2\lambda_4\,, \ \lambda_4 < 0 \label{eq: alignment selection Delta 54}
\end{eqnarray}

The potential is CP conserving, so the alignments $A$ and $A'$ merge. The alignments $B$ and $C$ remain distinct in the $\Delta(54)$ case, as they are not related by any symmetry transformation.

\subsection{The physical Higgs bosons}

Since the group representation are triplets of $\Delta(54)$ and doublets of $SU(2)$ there are 12 fields. After the Spontaneous Symmetry Breaking 3 of the fields give mass to the $W^\pm$ and $Z$ bosons and the number of fields get reduced to 9. In these 9 fields 5 are neutral and 4 are charged.
In this work, we are interested in the masses of the Higgs bosons at alignment $C$. The masses are as follows:

\begin{eqnarray}
m_{h_{SM}}^2 &=& 2m^2\,, \nonumber\\[1mm]
m_{H^\pm}^2 &=& v^2 (\lambda_{2} - 2 \lambda_{4})\quad \mbox{(double degenerate)}\,,\nonumber\\[1mm]
m_{h}^2 &=& \frac{v^2}{3} (3 \lambda_{3} - 7 \lambda_{4})\quad \mbox{(double degenerate)}\,,\nonumber\\[1mm]
m_{H}^2 &=& 3 v^2 (\lambda_{3} - \lambda_{4})\quad \mbox{(double degenerate)}\,.\label{Delta 54-masses C}
\end{eqnarray}

At alignment $B$ the masses are given by:

\begin{eqnarray}
m_{h_{SM}}^2 &=& 2m^2\,, \nonumber\\[1mm]
m_{H^\pm}^2 &=& v^2 \lambda_{2}\quad \mbox{(double degenerate)}\,,\nonumber\\[1mm]
m_{h}^2 &=& v^2 (\lambda_{3} + \lambda_{4})\quad \mbox{(double degenerate)}\,,\nonumber\\[1mm]
m_{H}^2 &=& v^2 (3\lambda_{3} - \lambda_{4})\quad \mbox{(double degenerate)}\,.\label{Delta 54-masses B}
\end{eqnarray}

The masses respective to alignments $A$ and $A'$ are:

\begin{eqnarray}
m_{h_{SM}}^2 &=& 2m^2\,, \nonumber\\[1mm]
m_{H^\pm}^2 &=& v^2 (\lambda_{2} - 3 \lambda_{3} + \lambda_{4})\quad \mbox{(double degenerate)}\,,\nonumber\\[1mm]
m_{h}^2 &=& \frac{v^2}{3} (-6 \lambda_{3} + 4 \lambda_{4} - \sqrt{9 \lambda_{3}^2 - 12 \lambda_{3}\lambda_{4} + 7 \lambda_{4}} )\quad \mbox{(double degenerate)}\,,\nonumber\\[1mm]
m_{H}^2 &=& \frac{v^2}{3} (-6 \lambda_{3} + 4 \lambda_{4} + \sqrt{9 \lambda_{3}^2 - 12 \lambda_{3}\lambda_{4} + 7 \lambda_{4}} )\quad \mbox{(double degenerate)}\,.\label{Delta 54-masses A}
\end{eqnarray}

\subsection{Softly broken potential}

Without loss of generality and for simplicity, we added one diagonal SBP. We choose the SBP $m_{22}^2$. Like the $\Sigma(36)$ case the VEV will have the following form:

\begin{equation}
(v,v,v)/\sqrt{3}\, \xlongrightarrow{m_{22}^2} \,(v,u,v)/\sqrt{3}\,.
\end{equation}

\begin{figure}[ht]
    \centering
    \includegraphics[width=0.6\textwidth]{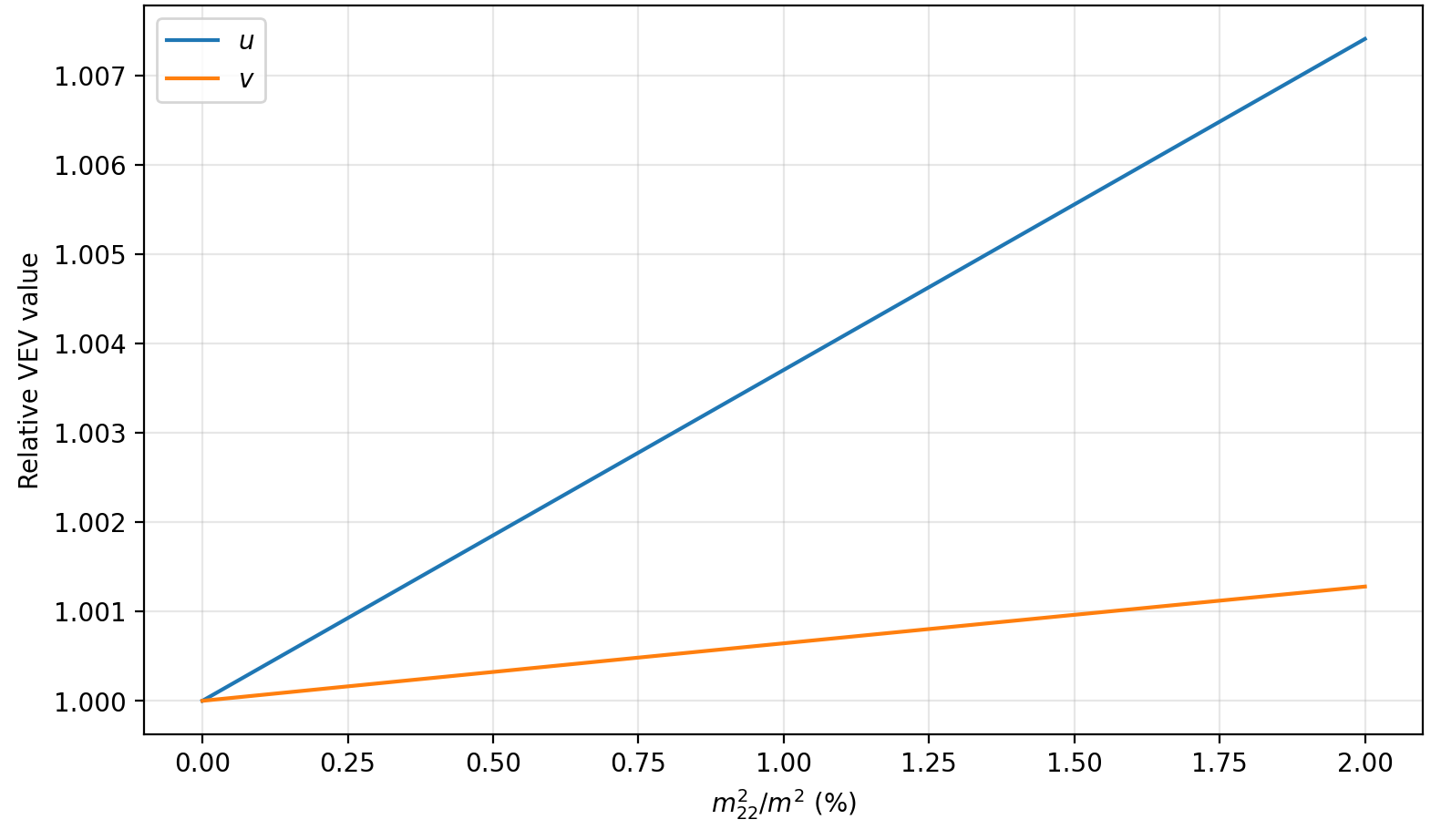}
    \caption{Evolution of the parameters $v$ and $u$ as a function of $m_{22}^2$ for $\Sigma(36)$.}
    \label{fig: uv Delta54}
\end{figure}

The evolution of $v$ and $u$ as a function of $m_{22}^2$ is presented in Figure \ref{fig: uv Delta54}.
The parameter $m_{22}^2$ was considered small compared to $m^2$.

The masses of the physical scalars are:

\begin{eqnarray}
m_{h_{SM}}^2 &=& -m^2 - \frac{m_{22}^2}{2} + \frac{4}{3} (u^2 + 2 v^2) \lambda_{1} + 
 \frac{u^2}{6} (\lambda_{3} + \lambda_{4}) - u v \lambda_{3} + u v \lambda_{4} + 
 \frac{v^2}{3} (4 \lambda_{3} + \lambda_{4}) \nonumber\\
 &&
 - \frac{1}{2} \sqrt{m_{22}^4 - 480 v^4 \lambda_{1} \lambda_{3} + 
\frac{u^4}{9} \lambda_{3}^2 - \frac{80}{3} u^2 v^2 \lambda_{3}^2 + 
\frac{824}{9} u v^3 \lambda_{3}^2 - 64 v^4 \lambda_{3}^2 + 
\frac{320}{3} u v^3 \lambda_{1} \lambda_{4}} \nonumber\\
&&
\overline{+ \frac{320}{3} v^4 \lambda_{1} \lambda_{4} + \frac{2}{9} u^4 \lambda_{3} \lambda_{4} - 4 u^2 v^2 \lambda_{3} \lambda_{4} - 
\frac{1000}{9} u v^3 \lambda_{3} \lambda_{4} + \frac{32}{9} v^4 \lambda_{3} \lambda_{4} + 
\frac{u^4}{9} \lambda_{4}^2 + \frac{68}{3} u^2 v^2 \lambda_{4}^2} \nonumber\\
&&
\overline{+ \frac{272}{3} u v^3 \lambda_{4}^2 + \frac{308}{9} v^4 \lambda_{4}^2 + 
\frac{4}{3} m^2 (2 v^2 (2 \lambda_{1} + 36 \lambda_{3} - 21 \lambda_{4}) + 
u^2 (2 \lambda_{1} - \lambda_{3} - \lambda_{4})} \nonumber\\
&&
\overline{- 2 u v (7 \lambda_{3} + 13 \lambda_{4})) - 
\frac{2}{3} m_{22}^2 (2 v^2 (-4 \lambda_{1} - 2 \lambda_{3} + \lambda_{4}) + 
u^2 (\lambda_{3} + \lambda_{4}) + 4 u v (20 \lambda_{1} + 7 \lambda_{3} + 13 \lambda_{4}))} \,, \nonumber\\
m_{H1^\pm}^2 &=& -m^2 - \frac{m_{22}^2}{2} + \frac{2}{3} u^2 \lambda_{1} + \frac{4}{3} v^2 \lambda_{1} + \frac{u^2}{6} \lambda_{2} + \frac{v^2}{3} \lambda_{2} - \frac{uv}{3} \lambda_{3} + \frac{v^2}{3} \lambda_{3} + \frac{uv}{3} \lambda_{4} \nonumber\\
&&
+ \frac{1}{2} \sqrt{m_{22}^4 + \frac{u^4}{9} \lambda_{2}^2 + \frac{8}{9} u^3 v \lambda_{2} \lambda_{4} + 
\frac{4}{9} v^4 (\lambda_{2}^2 + 3 \lambda_{3}^2 - 2 \lambda_{2} \lambda_{4} - 
4 \lambda_{3} \lambda_{4} + 2 \lambda_{4}^2)} \nonumber\\
&&
\overline{+ \frac{8}{9} u v^3 (3 \lambda_{2} \lambda_{3} - 3 \lambda_{3}^2 - 
4 \lambda_{2} \lambda_{4} + \lambda_{3} \lambda_{4} + 2 \lambda_{4}^2) + 
\frac{2}{3} m_{22}^2 (u^2 \lambda_{2} + 2 v^2 (-\lambda_{2} + \lambda_{3})} \nonumber\\
&&
\overline{+ 2 u v (\lambda_{2} - \lambda_{3} + \lambda_{4})) + 
\frac{4}{9} u^2 v^2 (\lambda_{2}^2 + 3 \lambda_{3}^2 + 2 \lambda_{3} \lambda_{4} + 
3 \lambda_{4}^2 - \lambda_{2} (6 \lambda_{3} + \lambda_{4}))}\,,\nonumber\\
m_{H2^\pm}^2 &=& -m^2 + \frac{u^2}{3} (2 \lambda_{1} + \lambda_{2}) + 
 \frac{2}{3} v^2 (2 \lambda_{1} + \lambda_{2} - \lambda_{3}) + 
 \frac{2}{3} u v (\lambda_{3} - \lambda_{4})
\,,\nonumber\\
m_{h1}^2 &=& -m^2 - \frac{m_{22}^2}{2} + \frac{4}{3} (u^2 + 2 v^2) \lambda_{1} + 
 \frac{u^2}{6} (\lambda_{3} + \lambda_{4}) - u v \lambda_{3} + u v \lambda_{4} + 
 \frac{v^2}{3} (4 \lambda_{3} + \lambda_{4}) \nonumber\\
 &&
 + \frac{1}{2} \sqrt{m_{22}^4 - 480 v^4 \lambda_{1} \lambda_{3} + 
\frac{u^4}{9} \lambda_{3}^2 - \frac{80}{3} u^2 v^2 \lambda_{3}^2 + 
\frac{824}{9} u v^3 \lambda_{3}^2 - 64 v^4 \lambda_{3}^2 + 
\frac{320}{3} u v^3 \lambda_{1} \lambda_{4}} \nonumber\\
&&
\overline{+ \frac{320}{3} v^4 \lambda_{1} \lambda_{4} + \frac{2}{9} u^4 \lambda_{3} \lambda_{4} - 4 u^2 v^2 \lambda_{3} \lambda_{4} - 
\frac{1000}{9} u v^3 \lambda_{3} \lambda_{4} + \frac{32}{9} v^4 \lambda_{3} \lambda_{4} + 
\frac{u^4}{9} \lambda_{4}^2 + \frac{68}{3} u^2 v^2 \lambda_{4}^2} \nonumber\\
&&
\overline{+ \frac{272}{3} u v^3 \lambda_{4}^2 + \frac{308}{9} v^4 \lambda_{4}^2 + 
\frac{4}{3} m^2 (2 v^2 (2 \lambda_{1} + 36 \lambda_{3} - 21 \lambda_{4}) + 
u^2 (2 \lambda_{1} - \lambda_{3} - \lambda_{4})} \nonumber\\
&&
\overline{- 2 u v (7 \lambda_{3} + 13 \lambda_{4})) - 
\frac{2}{3} m_{22}^2 (2 v^2 (-4 \lambda_{1} - 2 \lambda_{3} + \lambda_{4}) + 
u^2 (\lambda_{3} + \lambda_{4}) + 4 u v (20 \lambda_{1} + 7 \lambda_{3} + 13 \lambda_{4}))} \,, \nonumber\\
m_{h2}^2 &=& m^2 + \frac{2}{3} v^2 (-2 \lambda_{1} + \lambda_{3}) + 
\frac{u^2}{3} (-2 \lambda_{1} - 3 \lambda_{3} + \lambda_{4}) + 
\frac{2}{3} u v (-\lambda_{3} + \lambda_{4})\,,\nonumber\\
m_{H1}^2 &=& -m^2 - \frac{m_{22}^2}{2} + \frac{2}{3}(u^2 + 2 v^2)\lambda_{1} + 
\frac{u^2}{6} (3 \lambda_{3} - \lambda_{4}) + \frac{u v}{3} (-\lambda_{3} + \lambda_{4}) + \frac{v^2}{3}(4\lambda_{3} - \lambda_{4})\nonumber\\
&&
+ \frac{1}{2} \sqrt{m_{22}^4 - \frac{32}{9} u^3 v \lambda_{4}^2 + \frac{u^4}{9} (3 \lambda_{3} - \lambda_{4})^2 + 
\frac{16}{9} u v^3 (3 \lambda_{3}^2 - 7 \lambda_{3} \lambda_{4} + 6 \lambda_{4}^2)}\nonumber\\
&&
\overline{- \frac{4}{9} u^2 v^2 (6 \lambda_{3}^2 - 17 \lambda_{3} \lambda_{4} + 13 \lambda_{4}^2) + 
\frac{4}{9} v^4 (12 \lambda_{3}^2 - 28 \lambda_{3} \lambda_{4} + 17 \lambda_{4}^2) + 
\frac{m_{22}^2}{3} (8 u v (\lambda_{3} - 3 \lambda_{4})}\nonumber\\
&&
\overline{+ u^2 (6 \lambda_{3} - 2 \lambda_{4}) + 4 v^2 (-2 \lambda_{3} + \lambda_{4}))} \,,\nonumber\\
m_{H2}^2 &=& -m^2 + \frac{2}{3} v^2 (2 \lambda_{1} + \lambda_{3}) + 
   2 u v (\lambda_{3} - \lambda_{4}) + 
   \frac{u^2}{3} (2 \lambda_{1} + \lambda_{3} + \lambda_{4})\,.
   \label{eq: Delta 54-masses}
\end{eqnarray}

It is simple to verify that by setting $m_{22}^2 = 0$ and $u = v$, one obtains the masses in eq.(\ref{Delta 54-masses C}).

The value of the potential at the VEV is given by:

\begin{equation}
    V|_{VEV} = \frac{1}{6} (-m^2 (u^2 + 2 v^2) + u (-m_{22}^2 u + \frac{4}{3} v^2 (u + 2 v) \lambda4))
\end{equation}

\subsection{Computational Example}

Just like the $\Delta(54)$ case we will choose a computational example in order to understand the results of the previous section. The parameters chosen gave us the following masses:

\begin{eqnarray}
    m_{h_{SM}} &=& 129.0 GeV\nonumber\\
    m_{H1^\pm} &=& 136.2 GeV\nonumber\\
    m_{H2^\pm} &=& 136.1 GeV\nonumber\\
    m_{h1} &=& 156.4 GeV\nonumber\\
    m_{h2} &=& 159.9 GeV\nonumber\\
    m_{H1} &=& 255.0 GeV\nonumber\\
    m_{H2} &=& 255.1 GeV\nonumber\\
\end{eqnarray}

\begin{figure}[ht]
    \centering
    \includegraphics[width=0.6\textwidth]{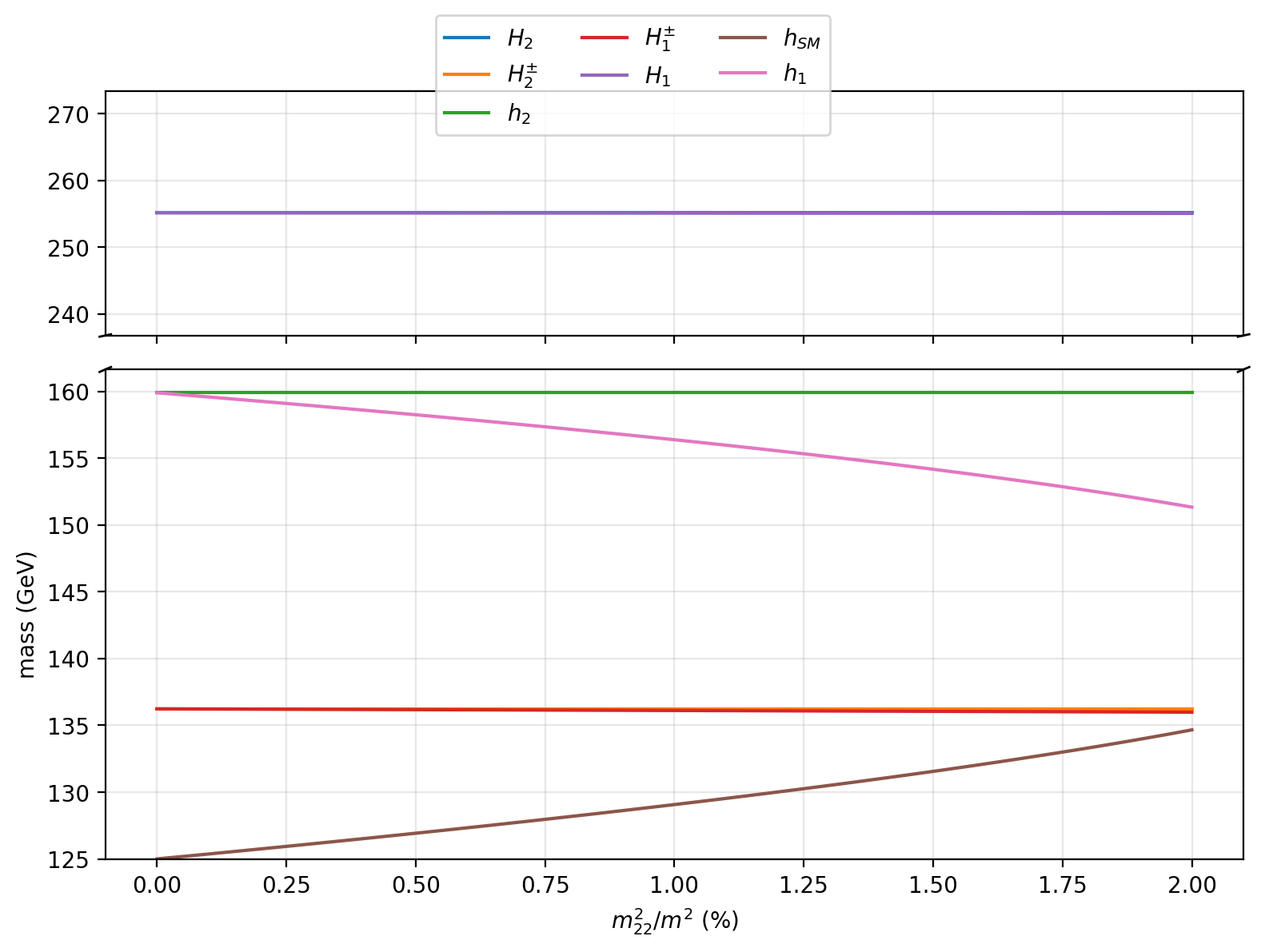}
    \caption{Masses evolution as a function of $m_{22}^2$ for $\Delta(54)$.}
    \label{fig: Masses Delta54}
\end{figure}

As expected, the soft breaking term lifted the degeneracies between several masses.
The evolution of the masses as a function of $m_{22}^2$ is shown in Figure \ref{fig: Masses Delta54}.

Just like the previous case, the charged Higgs bosons split into two distinct pairs, the light and heavy neutral scalars pairs masses also get split into four distinct scalars.
The magnitude of the splitting varies between states.

All the cases with a single SBP turned on were systematically examined, but the results are similar to those shown previously so we do not present them explicitly.


\section{Decays and residual symmetries}\label{section: decays}

In the symmetric limit without SBPs, each possible alignment preserves some residual symmetry, and therefore we expect a scalar sector with stable states. When the symmetry is softly broken, this is generally not the case, but there can be special cases for specific combinations of alignments and SBPs \cite{deMedeirosVarzielas:2022kbj}.

Consider the case where the SBP $m_{22}^2$ is turned on and the VEV alignment changes from the $(1,1,1)$ direction and is given by $(v, u, v)/\sqrt{3}$. The changed alignment no longer preserves a $\Z_3$ residual symmetry, but a residual $\Z_2$ symmetry remains as a symmetry of the potential. The generator $g$ of this $\Z_2$ group has the following actions $\phi_{1} \, \xrightarrow{g} \, \phi_{3}$ and $\phi_{3} \, \xrightarrow{g} \, \phi_{1}$. So $g$ has the following action on the physical states: $h_{SM} \, \xrightarrow{g} \, h_{SM}$, $h_{1} \, \xrightarrow{g} \, h_{1}$ and $h_{2} \, \xrightarrow{g} \, -h_{2}$. It then follows that only certain decays are allowed at all loop levels. Decays of $h_{2}$ into $h_{SM/1}$ are not allowed, as vertices $h_{SM/1}, h_{SM/1}, h_{2}$ and $h_{2}, h_{2}, h_{2}$ are forbidden by the residual symmetry.

In this case, $h_{2}$ is a potential dark matter candidate. Before $m_{22}^2$ is turned on there are two degenerate light bosons, but after $m_{22}^2$ is turned on these bosons masses get split and the $h_{2}$ is the heavier one. The eigenvectors associated with each scalar in the eq.(\ref{eq: Sigma36-generators}) and eq.(\ref{eq: Delta 54-generators}) basis have the following form:

\begin{equation}
n_{h_{SM}} = \frac{1}{\sqrt{2+a^2}}\triplet{1}{a}{1}\,,\quad
n_{h_{1}} = \frac{1}{\sqrt{2+b^2}}\triplet{1}{b}{1}\,,\quad 
n_{h_{2}} = \frac{1}{\sqrt{2}}\triplet{-1}{0}{1}\,.\label{eq: eigenvectors}
\end{equation}

Where the expressions for $a$ and $b$ depend on the symmetry in study. We have found analytic expressions for these quantities, but they are cumbersome. In the exact case (where $m_{22}^2 = 0$ and $u = v$) these quantities acquire the values $a = 1$ and $b = -2$.

\begin{figure}[ht]
    \centering
    \includegraphics[width=0.6\textwidth]{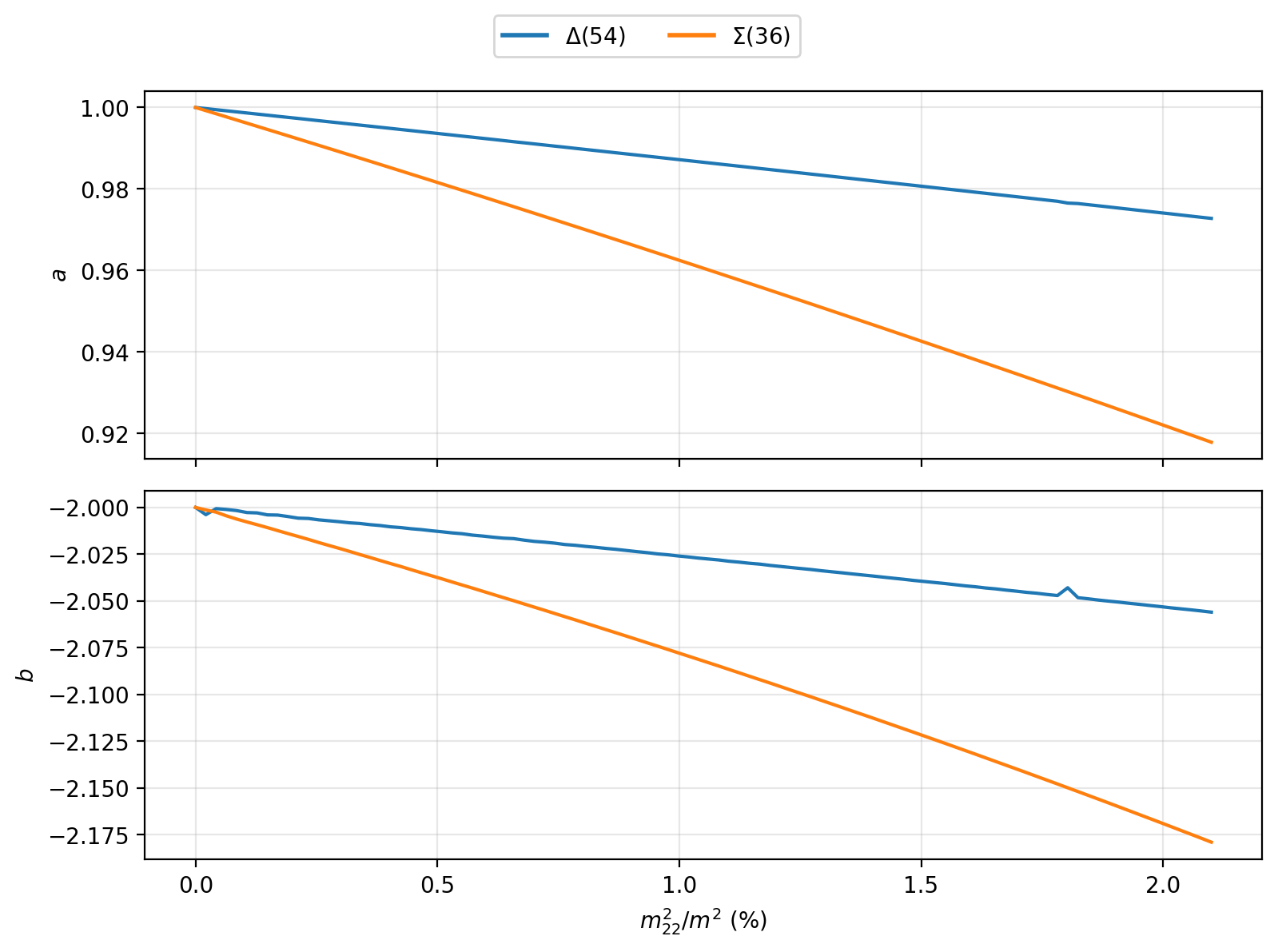}
    \caption{Evolution of the parameters $a$ and $b$ as a function of $m_{22}^2$ for $\Sigma(36)$ and $\Delta(54)$.}
    \label{fig: abParameters}
\end{figure}

To better understand the evolution of these parameters with the SBP $m_{22}^2$, the Figure \ref{fig: abParameters} was generated. 
Increasing the relative magnitude of $m_{22}^2$ makes the parameters $a$ and $b$ deviate further from their original values.

Since the eigenvectors associated with each state must be orthogonal, $a$ and $b$ are related by $a = -2/b$.

Consider now the qualitatively similar case where the SBPs $m_{22}^2$ and $m_{13}^2$ are both turned on. In this situation, the same $\Z_2$ still remains as a residual symmetry. Although the expressions for the masses and eigenvectors now depend on the two SBPs rather than one, they are still of the same shape (differentiating the second component of the triplet, as seen in eq.(\ref{eq: eigenvectors})). $h_2$ is still stabilized against decays and is a dark matter candidate.

In the most general case with more SBPs turned on, the residual symmetry is not present and all decays would be allowed.

\section{Discussion and conclusions}\label{section: conclusion}

We analysed the scalar sector of 3 Higgs Doublet Models with $\Delta(54)$ and $\Sigma(36)$ symmetries, that are softly broken with terms that will change the direction of the vacuum alignments with respect to the exact symmetry limit.

We systematically considered each of the possible vacuum expectation alignments and each of the soft breaking terms, studying the effect on the alignments. We focused on the qualitatively different situations, with a single soft breaking term active, particularly as these are the cases where residual symmetries survive.

We checked how the physical mass eigenstates and respective masses change with the soft breaking parameters. As expected, the mass degeneracies of the symmetric limit are generically lifted. In some of the cases the analytical expressions are relatively brief and were presented, and some numerical examples were also presented to more clearly show the lifting of the mass degeneracies.

The decays of the physical states were analysed, and we found and highlight cases where a residual unbroken symmetry stabilizes a possible scalar dark matter candidate against decays - this is a situation that had been found previously in an $A_4$ invariant scalar potential with a soft breaking term that preserve the respective vacuum alignment, but in these potentials it appears instead with a soft breaking term that changes the vacuum alignment.


\section*{Acknowledgments}
IdMV thanks the University of Basel for hospitality.
IdMV acknowledges funding from Fundação para a Ciência e a Tecnologia (FCT) through the FCT Mobility program, and through
the projects CFTP-FCT Unit 
UID/00777/2025 (\url{https://doi.org/10.54499/UID/00777/2025}),  UIDB/FIS/00777/2020 and UIDP/FIS/00777/2020, CERN/FIS-PAR/0019/2021,
CERN/FIS-PAR/0002/2021, 2024.02004 CERN, which are partially funded through POCTI (FEDER), COMPETE,
QREN and EU.


\end{document}